\documentstyle[11pt]{article}
\def\fnote#1#2{\begingroup\def\thefootnote{#1}\footnote{#2}\addtocounter
{footnote}{-1}\endgroup}

\begin{document}

\hfill{UTTG-18-11}

\vspace{36pt}

\begin{center}
{\large {\bf {Collapse of the State Vector}}}

\vspace{36pt}
Steven Weinberg\fnote{*}{Electronic address:
weinberg@physics.utexas.edu}\\
{\em Theory Group, Department of Physics, University of
Texas\\
Austin, TX, 78712}

\vspace{30pt}

\noindent
{\bf Abstract}
\end{center}

\noindent
Modifications of quantum mechanics are considered, in which the state vector of any system, large or small, undergoes a stochastic evolution.  The general class of theories is described, in which the probability distribution of the state vector collapses to a sum of delta functions, one for each possible final state, with coefficients given by the Born rule.    

\vfill

\pagebreak

\begin{center}
{\bf I. INTRODUCTION}
\end{center}

There is now in my opinion no entirely satisfactory interpretation of quantum mechanics[1].  The Copenhagen interpretation[2] assumes a mysterious division between the microscopic world governed by quantum mechanics and a macroscopic world of apparatus and observers that obeys classical physics.  During measurement the state vector of the microscopic system collapses in a probabilistic way to one of a number of classical states, in a way that is unexplained, and cannot be described by the time-dependent Schr\"{o}dinger equation.  The many-worlds interpretation[3]  assumes that the state vector of the whole of any isolated system does not collapse, but evolves deterministically according to the time-dependent Schr\"{o}dinger equation.  In such a deterministic theory it is hard to see how probabilities can arise.  Also, the branching of  the world into vast numbers of histories  is  disturbing, to say the least.  The decoherent histories approach[4] like the Copenhagen interpretation gives up on the idea that it is possible to completely characterize the state of an isolated system at any time by a vector in Hilbert space, or by anything else, and instead provides only a set of rules for calculating the probabilities of certain kinds of history.  This avoids inconsistencies, but without any objective characterization of the state of a system, one wonders where the rules come from.

Faced with these perplexities, one is led to consider the possibility that quantum mechanics needs correction.  There may be a Hilbert space vector that completely characterizes the state of a system, but that suffers an inherently probabilistic physical collapse, not limited as in the Copenhagen interpretation to measurement by a macroscopic apparatus, but occurring at all scales, though presumably much faster for large systems.    From time to time specific models for this sort of collapse have been proposed[5].  In the present  article we will consider the properties of theories of the stochastic evolution of the state vector in a more general formalism.  We  assume that this evolution depends only on the state vector, with no hidden variables.  In contrast to earlier work, we concentrate on the linear first-order differential equation that in general describes the evolution of the probability distribution of the state vector in Hilbert space.  We find conditions on this evolution so that it leads to final states with probabilities given by the Born rule of ordinary quantum mechanics.  This general formalism is also applied to the special case of a state vector that evolves through quantum jumps.  Theories of the evolution of the density matrix are examined as an important  special case of the more general formalism.

\begin{center}
{\bf II. EVOLUTION OF THE STATE VECTOR'S PROBABILITY DENSITY}
\end{center}

We consider a general isolated system, which may or may not include a macroscopic measuring apparatus and/or an observer.  We assume as in ordinary quantum mechanics that the state of the system is entirely described by a vector in Hilbert space.  The state vector here is taken in a sort of Heisenberg picture, in which operators $A(t)$ have a time dependence dictated by the Hamiltonian $H$ as $\exp(iHt)A(0)\exp(-iHt)$.  But the state vector in this sort of theory is not time-independent; it  undergoes  a stochastic evolution,  slow for microscopic systems but rapid for larger systems, so that at any time $t$ there is a probability  $P(\psi,t)d\psi$ for the wave function to be in a small volume $d\psi$ around any value $\psi$.  Here we are  adopting a basis that is so far arbitrary, labeled by a discrete index $i$, so that $\psi$ is an abbreviation for the whole set of components $\psi_i$ and $\psi_i^*$, constrained by the normalization condition $\sum_i|\psi_i|^2=1$, and  $d\psi$ is defined as 
\begin{equation}
d\psi\equiv \delta\left(1-\sum_i|\psi_i|^2\right)\prod_i d|\psi_i|^2\;\frac{d{\rm Arg}\psi_i}{2\pi}\;,
\end{equation}
a measure invariant under unitary transformations of the $\psi_i$.
The continuum case will be considered later, in Sec. IV.

We assume time-translation invariance, so that if the wave function at time $t$ has a definite value $\psi$, then at a later time $t'$ the probability density at $\psi'$ will be some function $\Pi(\psi',\psi,t'-t)$ of $\psi'$, of $\psi$, and of the elapsed time $t'-t$, but not separately of $t$ or $t'$.   It follows then from the rules of probability that if  at time $t$ the wave function has a probability density $P(\psi,t)$, then at time $t'$ the probability density will be
\begin{equation}
P(\psi',t')=\int d\psi\;\Pi(\psi',\psi,t'-t)\,P(\psi,t)\;.
\end{equation}
Differentiating with respect to $t'$ and then setting $t'=t$ gives our fundamental differential equation for the evolution of the probability density:
\begin{equation}
\frac{d}{dt}P(\psi',t)=\int d\psi\;K_{\psi',\psi}\,P(\psi,t)\;,
\end{equation}
where $K$ is the kernel 
\begin{equation}
K_{\psi',\psi}\equiv \left[\frac{d}{d\tau}\Pi(\psi',\psi,\tau)\right]_{\tau=0}\;,
\end{equation}  which depends on the details of the system under study, including any measuring apparatus that the system may contain.    Eq.~(3) resembles the time-dependent Schr\"{o}dinger equation with $K$ in place of $-iH$, because both follow from time-translation invariance, but Eq.~(3) describes the evolution of the probability density in Hilbert space rather than of the state vector, and so $K$ is real rather than anti-Hermitian.  Like the time-dependent Schr\"{o}dinger equation, Eq.~(3) neither violates nor guarantees Lorentz invariance.  Presumably, in a Lorentz invariant theory, $K$ would be accompanied with other kernels that describe how probabilities change with the position of the observer.

  The solution of Eq.~(3) is of course
\begin{equation}
P(\psi',t)=\int d\psi \left(e^{Kt}\right)_{\psi',\psi}P(\psi,0)\;,
\end{equation}
with the exponential of $Kt$ defined as usual by its power series expansion.  
To evaluate this exponential, we let $f_N(\psi)$ be the linearly independent right-eigenfunctions of $K$:
\begin{equation}
\int d\psi\;K_{\psi',\psi}\,f_N(\psi)=-\lambda_N f_N(\psi')\;,
\end{equation}
with eigenvalues $-\lambda_N$.  Because there is no need for $K$ to be Hermitian, some of the eigenfunctions and eigenvalues may be complex, but because $K$ is real, any complex eigenfunctions and eigenvalues must come in complex conjugate pairs.

We will assume that the $f_N(\psi)$ form a complete set.  This is the generic case; other cases can be handled by letting some eigenvalues and eigenfunctions of $K$ merge with each other.  Where the $f_N$  form a complete set we may write the kernel as
\begin{equation}
K_{\psi',\psi}= -\sum_N \lambda_N\,f_N(\psi')\,g_N(\psi)\;,
\end{equation}
where  $g_N(\psi)$ are some coefficient functions, not related in any simple way to $f_N(\psi)$.  The eigenvalue condition (6) requires that
\begin{equation}
\int d\psi\; g_M(\psi)\,f_N(\psi)=\delta_{NM}\;.
\end{equation}
Then $g_N$ will  be a left-eigenfunction of $K$, also with eigenvalue $-\lambda_N$:
\begin{equation}
\int d\psi'\;g_N(\psi')\,K_{\psi',\psi}=-\lambda_N\, g_N(\psi)\;.
\end{equation}
(Eq.~(7) does not define $g_N$ in the case $\lambda_N=0$; in this case the definition is provided by Eqs.~(8) and (9).)  The completeness relation for the $f_N$ can then be expressed as
\begin{equation}
1_{\psi',\psi}=\sum_N f_N(\psi')\,g_N(\psi)
\end{equation}
where $1_{\psi',\psi}$ is defined so that, for any smooth function $F(\psi)$,
\begin{equation}
\int d\psi\;1_{\psi',\psi}\,F(\psi)=F(\psi')\;.
\end{equation}
It is elementary then to use the power series expansion for the exponential to calculate that
\begin{equation}
\left[e^{Kt}\right]_{\psi',\psi}=\sum_N e^{-\lambda_Nt}f_N(\psi')\,g_N(\psi)\;.
\end{equation}
  The probability distribution for the wave function is therefore
\begin{equation}
P(\psi,t)=\sum_N e^{-\lambda_Nt}f_N(\psi)\,\int d\psi'\;g_N(\psi')\,P(\psi',0)\;.
\end{equation}
(Where the $f_N$ miss being a complete set by a finite number of terms, the exponentials are in general accompanied with polynomial functions of time.)

\begin{center}
{\bf III. LIMIT OF EVOLUTION}
\end{center}

It is  clear that in order for the probability distribution to approach any sort of limit   for $t\rightarrow \infty$, all the   eigenvalues must have negative real parts; that is, ${\rm Re}\lambda_N\geq 0$.  If we assume that there is a minimum value to the smallest non-zero value of ${\rm Re}\lambda_N$, then  the probability distribution becomes dominated by the zero modes:  for  $t\rightarrow \infty$
\begin{equation}
P(\psi,t)\rightarrow \sum_n f_n(\psi)\,\int d\psi'\;g_n(\psi')\,P(\psi',0)\;,
\end{equation}
where $n$ runs over the values of $N$ for which $ \lambda_N=0$.  (The contribution of eigenmodes with ${\rm Re}\lambda_N=0$ but  ${\rm Im}\lambda_N\neq 0$ presumably oscillates so rapidly as $t\rightarrow \infty$ as to be unobservable.)  The $f_n(\psi)$ can be regarded as fixed points of the differential equation (3).
  The magnitude of the non-zero eigenvalues depends on the nature of the system in question.  Presumably where a system is large, as in measurement by a macroscopic apparatus, the  values of the non-zero eigenvalues are large, in which case the approach to the limit (14) is exponentially fast.

Although the limit of the probability distribution for $t\rightarrow \infty$ depends only on the zero-modes
$f_n$ and $g_n$, in general to calculate the evolution of the probability distribution for finite times we need to know all the eigenfunctions $f_N$ and $g_N$.  But the whole time dependence of the probability distribution can be calculated in terms of the zero modes in the special case                in which          all non-zero $\lambda_N$ are equal, say to $\lambda$.  
Then Eq.~(12) gives
$$
\left[e^{Kt}\right]_{\psi',\psi}=\sum_n f_n(\psi')\,g_n(\psi)+e^{-\lambda t}\sum_\nu f_\nu(\psi')\,g_\nu(\psi)\;,
$$
where $ \nu$ runs over the values of $N$ for which $\lambda_N\neq 0$.  
The completeness relation (10) gives
$$
\sum_\nu f_\nu(\psi')\,g_\nu(\psi)=1_{\psi',\psi}-\sum_n f_n(\psi')\,g_n(\psi)
$$
so 
$$
\left[e^{Kt}\right]_{\psi',\psi}=\left[1-e^{-\lambda t}\right]\sum_n f_n(\psi')\,g_n(\psi)+e^{-\lambda t}\left[1\right]_{\psi',\psi}\;, $$
and the probability distribution is
\begin{equation}
 P(\psi,t)=P(\psi,0)e^{-\lambda t}+\left[1-e^{-\lambda t}\right]\sum_n f_n(\psi)\int d\psi'\,g_n(\psi')\,P(\psi',0) \;,
\end{equation}
in which we can see explicitly how  the probability distribution approaches the limit (14) for $t\rightarrow \infty$.

The kernel $K$ (including the zero modes $ f_n$ and $g_n$ along with the non-zero eigenvalues $-\lambda_\nu$) depends on the details of the system in question, as well as depending  on the as yet mysterious dynamics of the collapse process.   Consider a  system containing a subsystem with a complete  set of commuting observables whose eigenvalues are labeled by an index $n$, and a measuring apparatus that through a unitary evolution of the whole system becomes entangled with the subsystem in such a way that, when the subsystem is in the $n$ th eigenstate of the observables, the apparatus is in a unique state.  It is convenient to perform a unitary transformation to a new basis,  in which $\varphi_n$ is the component of the state vector along such  a joint state of the whole system.  In order to reproduce the results of the Copenhagen interpretation the probability distribution at  late times must relax to a sum over 
$n$ of terms proportional to $\prod_{m\neq n}\delta (|\varphi_m|^2)$, so that only $\varphi_n$ is allowed to be non-zero in the $n$th term.  To reproduce the Born rule, the coefficient of the $n$th term must be proportional to the initial value of $|\varphi_n|^2$.  Comparing with Eq.~(14), we see that the zero modes here can be labeled with the same index $n$, with
\begin{equation}
f_n(\varphi)={\cal F}_n({\rm Arg}\varphi_n)\prod_{m\neq n}\delta (|\varphi_m|^2)\;,~~~~g_n(\varphi)=|\varphi_n|^2\;,
\end{equation}
where ${\cal F}_n(\theta)$ is an unknown function satisfying $\int_0^{2\pi} {\cal F}_n(\theta)\,d\theta=2\pi$.  
The normalization of these zero modes has been chosen to be consistent with Eq.~(8), which requires that $\int d\varphi f_n(\varphi)g_m(\varphi)=\delta_{nm}$.  Also, since 
it is only $f_ng_n$ that enter in this requirement, we have made an arbitrary choice of  a convenient normalization for $f_n$, thus fixing the normalization of $ g_n$.  According to Eq.~(14), the probability density at late times becomes
\begin{equation}
P(\varphi,t)\rightarrow \sum_n{\cal F}_n({\rm Arg}\varphi_n)\left(\prod_{m\neq n}\delta\Big(|\varphi_m|^2\Big)\right)\,\int d\varphi'|\varphi_n'|^2\,P(\varphi',0) \;.
\end{equation}  Note that here Eqs.~(9) and (16) give, for each $n$ and $\varphi$
\begin{equation}
\int d\varphi'\;|\varphi'_n|^2 K_{\varphi',\varphi}=0\;.
\end{equation}
This implies the time-independence of the quantity
\begin{equation}
P_n\equiv \int d\varphi\;|\varphi_n|^2\,P(\varphi,t)\;.
\end{equation}
This makes sense, because $P_n$ according to the Born rule is the probability that, when the collapse is finished, the state of the system will be found in the basis state $n$, and this of course must be independent of  $t$.  
Since $\sum_n |\varphi'_n|^2=1$, the sum of Eq.~(18) over $n$ yields
\begin{equation}
\int d\varphi'\; K_{\varphi',\varphi}=0\;,
\end{equation}
which is the condition that Eq.~(3) respects the conservation of the total probability $\int d\varphi\,P(\varphi,t)$.

In usual measurements, the measuring apparatus does not evolve into a unique state when the subsystem is in the $n$th eigenstate of a set of observables, but into any one of a number of apparatus states, labeled with another index $r$.  It is convenient again to choose a corresponding basis, so that the components of the wave function are labeled $\varphi_{nr}$, with $\sum_{nr}|\varphi_{nr}|^2=1$.  In this case, assuming all apparatus states $r$ for a given subsystem state $n$ are equally probable,  consistency with the results of the Copenhagen interpretation and the condition $\int d\varphi\,f_ng_m=\delta_{nm}$ requires that
\begin{equation}
f_n(\varphi)= {\cal F}_n\prod_{r,m\neq n}\delta\left(|\varphi_{mr}|^2\right)\;,~~~~~g_n(\varphi)=\sum_r |\varphi_{nr}|^2\;,
\end{equation}
where ${\cal F}_n$ is an unknown function of the phases of all $\varphi_{nr}$, whose average over phases is unity, and the individual normalization of $f_n$ and $g_n$ has again been chosen for convenience.  The individual probabilities $\int g_n(\varphi)P(\varphi)d\varphi$ and the total probability $\int P(\varphi)d\varphi$ are conserved here for the same reason as before.

From the point of view adopted here, there is nothing special about measurement.  Measurement is just a process in which the state vector of a system (typically microscopic) becomes entangled with the state vector of a relatively large system, which then undergoes a collapse to an eigenstate of some operators determined by the characteristics of that system.  So we expect that the state vector of any system undergoes a similar collapse, but one that is much faster for large systems.    But collapse to what?  Without attempting a precise general prescription, we have in mind that these are the sorts of states familiar in classical physics.  For instance, in a Stern-Gerlach experiment,  they would be states in which a macroscopic detector registers that an atom has a definite trajectory, not a superposition of trajectories.  In Schr\"{o}dinger's macabre thought experiment[6], they are states in which the cat is alive, or dead, but not a superposition of alive and dead.  These states are like the ``pointer states'' of Zurek[7], but here these basis states are determined by the physics of the assumed collapse of the state vector, rather than by the decoherence produced by interaction with small external perturbations.

\begin{center}
{\bf IV. CONTINUUM STATES}
\end{center}

It is straightforward to adapt this formalism to the continuum case, where the wave functions depend on a continuous variable $x$ rather than a discrete label $i$.  In the continuum case, we take $\psi$ as an abbreviation for the functions $\psi(x)$ and $\psi^*(x)$, normalized so that $\int dx |\psi(x)|^2=1$;  the probability distribution $P[\psi,t]$ and the kernel $K_{\psi,\psi'}$ are functionals of these functions;  and $\int d\psi$ is a functional integral, with a normalization that can be chosen as convenience dictates.  There is no reason here to expect a gap between the zero and non-zero eigenvalues of $K$, and in the example discussed in Sec. VII there is no such gap, so we will not here bother to separate the zero-modes from the eigenfunctionals of $K$ with non-zero eigenvalue.  The kernel can  be expressed as
\begin{equation}
K_{\psi',\psi}=-\int dN \;\lambda _N\,f_N[\psi']g_N[\psi]\;,
\end{equation}
where
\begin{equation}
\int d\psi\; K_{\psi',\psi}f_N[\psi]= -\lambda_N f_N(\psi')\;,~~~\int d\psi\;g_{N'}[\psi]f_N[\psi] =\delta(N'-N)\;.
\end{equation} 
Using the completeness relation
\begin{equation}
1_{\psi',\psi}=\int dN\;f_N[\psi']g_N[\psi]\\;,
\end{equation}
we have
\begin{equation}
\left[e^{Kt}\right]_{\psi',\psi} =\int dN\;f_N[\psi']g_N[\psi]e^{-\lambda_N t}
\end{equation}
and the probability distribution at time $t$ is
\begin{equation}
P[\psi,t]=\int dN\; f_N[\psi]e^{-\lambda_N t}\int d\psi'\;g_N[\psi']P[\psi',0]\;.
\end{equation}
As before, to avoid runaway solutions we need to assume that ${\rm Re}\lambda_N\geq 0$ for all eigenvalues, in which case with increasing time Eq.~(26) is increasingly dominated by the eigenmodes with smallest $\lambda_N$.  But without a gap between  zero and non-zero eigenvalues, the probability distribution may not approach any specific limit exponentially as $t\rightarrow \infty$.

\begin{center}
{\bf V. QUANTUM JUMPS}
\end{center}

Our discussion so far has been very general, not dependent on any specific picture of the evolution of the state vector.  We can be a little more concrete, by assuming that the wave function undergoes a series of quantum jumps, from $\psi$ to $J\psi$, where $J$ is a non-linear operator depending on one or more random parameters.  If the rate of jumps is $\Gamma$, and the wave function at some time $t$ is $\psi'$, then at a slightly later time $t+dt$ the probability distribution at $\psi$ is
$(1-\Gamma dt)1_{\psi,\psi'}+\Gamma dt \left\langle 1_{\psi,J\psi'}\right\rangle$, where brackets indicate an average over the random parameters on which the operator $J$ depends.  Hence the evolution of the probability distribution is given by Eq.~(3), with kernel
\begin{equation}
K_{\psi,\psi'}=-\Gamma\left(1_{\psi,\psi'}-\left\langle 1_{\psi,J\psi'}\right\rangle\right)\;.
\end{equation}
We note in particular that, for any function (or functional) $g(\psi)$ of the wave function, we have
\begin{equation}
\int d\psi\;g(\psi)\,K_{\psi,\psi'}=-\Gamma\Big(g(\psi')-\left\langle g(J\psi')\right\rangle\Big)\;,
\end{equation}
so the condition for $g(\psi)$ to be a left eigenfunction of the kernel is that
\begin{equation}
\left\langle g(J\psi)\right\rangle=\Lambda\,g(\psi)\;,
\end{equation}
in which case the corresponding eigenvalue is
\begin{equation}
\lambda =-\Gamma\,(1-\Lambda)\;.
\end{equation}
The left-eigenfunctions of the kernel with zero eigenvalue are those functions $g(\psi)$ that {\em on average} are unaffected by quantum jumps.

\begin{center}
{\bf VI. DENSITY MATRIX}
\end{center}

The class of theories presented here are more general than in any based on an assumed differential equation for the density matrix, as there is much more information contained in the probability distribution $P(\psi)$ than in the density matrix.  (For instance, for a system with two discrete states, the density matrix is specified by only three real parameters, while the probability distribution is an unknown real {\em function} of one modulus and two phases.) The density matrix in is defined in a general discrete basis by
\begin{equation}
\rho_{ij}(t)\equiv \int d\psi\;P(\psi,t)\psi_i\psi_j^*\;.
\end{equation}
In particular, Eq.~(3) gives the rate of change of the density matrix 
\begin{equation}
\frac{d}{dt}\rho_{ij}(t)\equiv \int d\psi\;\int d\psi'\;K_{\psi,\psi'}P(\psi',t)\,\psi_i\psi^*_j\;.
\end{equation}
In order for the right-hand side to be expressible in terms of  $\rho$, we would need the space of bilinear functions of $\psi$ to be invariant under the left action of the kernel $K$:
\begin{equation}
\int d\psi\;K_{\psi,\psi'}\,\psi_i\psi_j^*=\sum_{i'j'}\;\kappa_{ij,i'j'}\,\psi'_{i'}\psi^{'*}_{j'}\;.
\end{equation}
This condition is preserved if we make a change of basis  by a unitary transformation of the wave function,  of course with a transformed $\kappa$ matrix.  Not all conceivable kernels satisfy a condition like Eq.~(33).  Where this condition  holds, the density matrix obeys the differential equation
\begin{equation}
\frac{d}{dt}\rho_{ij}(t)=\sum_{i'j'}\;\kappa_{ij,i'j'}\,\rho_{i'j'}(t)\;.
\end{equation}

There are reasons to suppose that this must be the case.  It is a familiar feature of quantum mechanics that different statistical ensembles of individual states can yield the same density matrix.  Gisin[8] has shown that for any two such ensembles of states of a given physical system that have the same density matrix $\rho$, it is always possible to invent a second isolated physical system that can be entangled with the first, in such a way that measurements in the second system can drive the first system to one or the other of the two ensembles with density matrix $\rho$.  This does not lead to any possibility of communication between the two systems, provided the density matrix contains all information concerning any possible observation of the first system, {\em and} provided
that the subsequent evolution of the density matrix depends only on the density matrix, not on the particular statistical ensemble it represents.  But if Eq.~(33) were not satisfied, then the evolution of the density matrix would depend on  the specific statistical ensemble of state vectors, not just on the density matrix, and instantaneous communication between isolated systems would be possible.

Where the probability distribution approaches the limit (17) at late time, in the basis $\varphi_n$ described in Section III, the density matrix becomes diagonal
\begin{equation}
\rho_{nm}\equiv \int d\varphi\;P(\varphi)\,\varphi_n\,\varphi_m^*\rightarrow P_n\delta_{nm}
\end{equation}
where the $P_n$ are constants given by Eq.~(19).  Of course, by a unitary transformation the density matrix can be put in a diagonal form at any time, but with diagonal elements and in a basis that change with time.   Eq.~(35) tells us that the density matrix approaches a diagonal form with in a fixed basis and with fixed diagonal elements, equal to the expectation values (19) of the density matrix at any time.

\begin{center}
{\bf VII. THE GRW CASE}
\end{center}

Finally, it is interesting to examine how  the theory proposed in the well-known paper of Ghirardi, Rimini, and Weber[5] (henceforth GRW) appears in the more general formalism presented here.  GRW suggested a stochastic evolution of the state vector, leading to its localization, and expressed their model in  a differential equation for the density matrix  for a single particle (written here using the Heisenberg picture described above, and in a  notation slightly different from that of GRW)
\begin{equation}
\frac{d}{dt}\rho_{x',x}(t)=-\omega\left(1-e^{-\alpha(x'-x)^2/2}\right)\rho_{x',x}(t)\;,
\end{equation}
with $\omega>0$ and $\alpha>0$.  (Here $x$ is the eigenvalue of the one-dimensional Heisenberg-picture position operator $\hat{x}(t)=\hat{x}(0)+\hat{p}t/m$.)  
Thus the  condition (33) here reads
\begin{equation}
\int d\psi\;K_{\psi,\psi'}\,\psi(x')\psi^*(x)=-\omega\left(1-e^{-\alpha(x'-x)^2/2}
\right)\,\psi'(x')\psi'^*(x)\;,
\end{equation}
so the kernel has eigenvalues
\begin{equation}-\lambda_{xx'}=-\omega\left(1-e^{-\alpha(x'-x)^2/2}\right)\leq 0
\;,\end{equation}
with left-eigenfunctionals
\begin{equation}
g_{xx'}[\psi]=\psi(x')\psi^*(x)\;.
\end{equation}
But this does not determine the kernel, since without changing Eq.~(37) we can change $K$ by adding any kernel $K^{(0)}$ for which
 \begin{equation}
\int d\psi\;K^{(0)}_{\psi,\psi'}\,\psi(x)\psi^*(x')=0\;.\end{equation}
The zero modes (among others that would depend on $K^{(0)}$) are the $g_{xx'}[\psi]$ given by Eq.~(39), with  $x=x'$.  This is a case where there is no gap between the negative eigenvalues and zero, and the probability distribution does not approach any definite limit, though the density matrix becomes increasingly diagonal as $t\rightarrow \infty$.  
	
Bell[5] subsequently gave  formulas for a jump operator $J$  and for the probability distribution for the random parameter in $J$ that would  yield the GRW equation (36) for the evolution of the density matrix.  (Bell's formulas do not follow uniquely from the GRW equation (36) for the evolution of the density matrix, but they do follow from other assumptions in the GRW paper.)  In the one-particle one-dimensional case Bell's results (in a somewhat different notation) gives
\begin{equation} [J_\xi\psi](x)=j(x-\xi)\psi(x)/R(\psi,\xi) \;,\end{equation}
where $j(x)=(2\alpha/\pi)^{1/4} \exp(-\alpha x^2)$; $\xi$ is a random parameter with probability density $R^2(\psi,\xi)$, and $R(\psi,\xi)$ is determined by the normalization condition on $J\psi$:
\begin{equation} R^2(\psi,\xi)=\int d^3x\; |j(x-\xi)\psi(x)|^2\;.\end{equation}
Using Eq.~(27), and setting the jump frequency $\Gamma$ equal to $\omega$, we can use Eq.~(41) to find the kernel $K$.  Because of the $\psi$-dependence of $R(\psi,\xi)$, it is not so easy here to find general solutions of  the eigenvalue condition (29).  But it is easy to see that Eq.~(29) is satisfied for the functionals $g_{x,x'}[\psi]\equiv \psi(x')\psi^*(x)$ for arbitrary $x$ and $x'$.  In these cases the factors $1/R$ in $[J_\xi\psi](x')$ and $[J_\xi\psi]^*(x)$ are  cancelled by the probability distribution $R^2$ for $ \xi$, and we find 
$\Lambda=\exp(-\alpha(x-x')^2/2)$.  Using Eq.~(30) then shows that $\psi(x')\psi^*(x)$ are left-eigenfunctionals of $K$ with eigenvalues (38), so Eq.~(37) is satisfied, and this yields the GRW equation (36).

I am grateful to James Hartle for several helpful conversations about this work and about the interpretation of quantum mechanics, and to Jacques Distler for valuable comments.  I also thank Angelo Bassi, Gian Carlo Ghirardi and Roderich Tumulka for  helpful correspondence about an earlier version of this paper.  This material is based upon work supported by the National Science Foundation under Grant Number PHY-0969020 and with support from The Robert A. Welch Foundation, Grant No. F-0014.

\vspace{10pt}

\begin{center}
{\bf ---------}
\end{center}

\vspace{10pt}

\begin{enumerate}
\item This point involves too many issues to be treated adequately  here.  The author's views on the present state of quantum mechanics are spelled out in detail in Section 3.7 of {\em Lectures on Quantum Mechanics}, (Cambridge University Press, 2012), to be published.
\item N. Bohr, Nature {\bf 121}, 580 (1928), reprinted in {\em Quantum Theory and Measurement}, eds. J. A. Wheeler and W. H. Zurek (Princeton University Press, Princeton, NJ, 1983); {\em Essays 1958--1962 on Atomic Physics and Human Knowledge} (Interscience, New York, 1963).
\item The published version is H. Everett, Rev. Mod. Phys. {\bf 29}, 454 (1957).
\item R. B. Griffiths, J. Stat. Phys. {\bf 36}, 219 (1984); R. Omn\`{e}s, Rev. Mod. Phys. {\bf 64}, 339 (1992); M. Gell--Mann and J. B. Hartle, in {\em Complexity, Entropy, and the Physics of Information}, ed. W. Zurek (Addison--Wesley, Reading, MA, 1990); in {\em Proceedings of the Third International Symposium on the Foundations of Quantum Mechanics in the Light of New Technology}, ed. S. Kobayashi, H. Ezawa, Y. Murayama, and S. Nomura (Physical Society of Japan, 1990); in {Proceedings of the 25th International Conference on High Energy Physics, Singapore, August 2--8, 1990}, ed. K. K. Phua and Y Yamaguchi (World Scientific, Singapore, 1990); J. B. Hartle, {\em Directions in Relativity, Vol. 1}, ed. B.-L. Hu, M.P. Ryan, and C.V. Vishveshwars (Cambridge University Press, Cambridge, 1993).  For a survey and more recent references, see P. Hohenberg, Rev. Mod. Phys. {\bf 82}, 2835 (2010).
\item See, e.g., D. Bohm and J. Bub, Rev. Mod. Phys. {\bf 38}, 453 (1966); P. Pearle, Phys. Rev. D {\bf 13}, 857 (1976); G. C. Ghirardi, A. Rimini, and T. Weber, Phys. Rev. D {\bf 34}, 470 (1986); J. S. Bell, in {\em Speakable and Unspeakable in Quantum Mechanics} (Cambridge University Press, Cambridge, 1987), pp. 201--212; L. Diosi, J. Phys. {\bf A21}, 2885 (1988); P. Pearle, Phys. Rev. A {\bf 39}, 2277 (1989); G. C. Ghirardi, P. Pearle, and A. Rimini, Phys. Rev. A {\bf 43}, 78 (1990); R. Penrose, in {\em Physics Meets Philosophy at the Planck Scale}, ed. C. Callender (Cambridge University Press, Cambridge, 2001), p. 290; S. Adler, D. C. Brody, T. A. Brun, and L. P. Hughston, J. Phys. A {\bf 34}, 8795 (2001).  For a review, see A. Bassi and G. C. Ghirardi, Phys. Rept. {\bf 379}, 257 (2003).  
\item E. Schr\"{o}dinger, Naturwiss. {\bf 23}, 807 (1935).
\item W. Zurek, Phys. Rev. D {\bf 24}, 1516 (1981).
\item N. Gisin, Helv. Phys. Acta {\bf 62}, 363 (1989).  The possibility of instantaneous communication between separated systems  is discussed in a wider context by J. Polchinski, Phys. Rev. Lett. {\bf 66}, 397 (1991).
\end{enumerate}

  \end{document}